\documentstyle{multi3}
\input{epsf.sty}
\begin{document}
\pagenumbering{roman}
\chapter{Interaction-Free Measurements}
\author[L.\ Vaidman]{Lev Vaidman}
\address{School of Physics and Astronomy,
Raymond and Beverly Sackler Faculty of Exact Sciences,
Tel-Aviv University, Tel-Aviv 69978, Israel}
\pagenumbering{arabic}

\section{The Penrose bomb testing problem}

I am  greatly indebted to Roger Penrose. I have learned very much from his
papers, from his exciting books, and from our (too short)
conversations.  I am most grateful to Roger for  developing  the idea of Avshalom Elitzur and myself on
interaction-free measurements (IFM). The version of IFM Penrose
described in his book (1994) is conceptually different from our
original proposal, and although it is much more difficult for
practical applications it has the advantage of demonstrating even more
striking quantum phenomena. So I will start with presenting Pen rose's
version of IFM.

Suppose we have a pile of bombs equipped with super-sensitive
triggers. The good bombs have a tiny mirror which is connected to a
detonator such that if any particle (photon) ``touches'' the mirror,
the mirror bounces and 
the bomb explodes.  Some of the bombs are duds in which the mirror is
rigidly connected to the massive body of the bomb.  Classically, the
only way to verify that a bomb is good is to touch the mirror, but
then a good bomb will explode. Our task is to test a bomb without
exploding it.  We are not allowed to make errors in our test, i.e., to
say that a bomb is good while it is a dud, but we
may sometimes cause an explosion.

There cannot be a solution by weighting the bomb, or touching the
mirror from the side, or any other similar way: the only observable physical
difference between a good bomb and a dud is that the good bomb will
explode when a single particle will touch the mirror, and the dud
will not.  Thus, the solution of this task seems to be logically
impossible: the only difference between the bombs is that one explodes
and the other does not and we are asked to test a bomb without
exploding it.  Nevertheless, quantum mechanics provides a solution to
the problem in a surprisingly simple way.

The method uses the well known  Mach-Zehnder interferometer which is
shown in  Fig. 1.   The photons reach the
first beam splitter which has transmission coefficient ${1/2}$.  The
transmitted and reflected parts of their waves are then reflected by
two mirrors and finally reunite at another similar beam splitter.
Two detectors collect the photons after they pass through the second
beam splitter.  It is possible to arrange the positions of the beam
splitters and the mirrors in such a way that due to destructive
interference no photons are detected by one of the detectors,
say $D_2$, and they all are detected by $D_1$.
\begin{figure}
\epsfysize=4.3 cm
 \centerline{\epsfbox{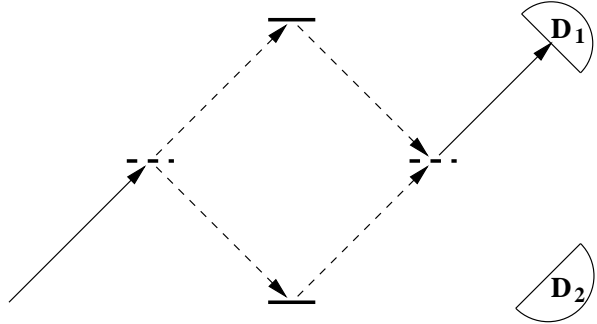}} 
\caption{{\bf Mach-Zehnder Interferometer.} When  the interferometer is
  properly tuned, all photons are detected by $D_1$ and none reach  $D_2$. The mirrors must be massive enough and have
  well-defined position.  }
\end{figure}

\begin{figure}[b]
\epsfysize=7 cm
 \centerline{\epsfbox{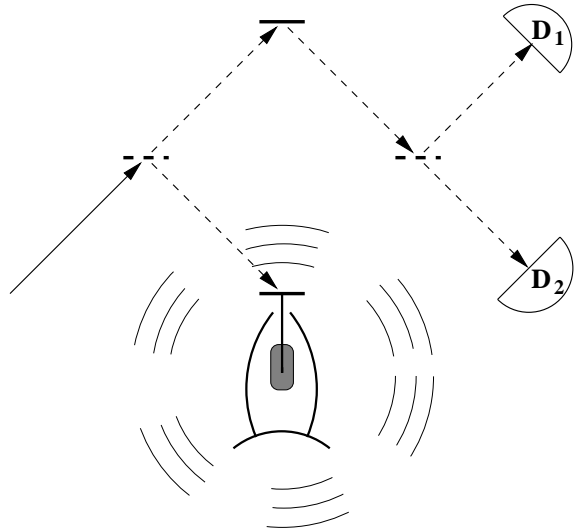}} 
\caption{{\bf The Penrose bomb-testing device.} The mirror of the good bomb
  cannot reflect the photon, since the incoming photon causes an
  explosion. Therefore, $D_2$ sometimes clicks. (The mirror of a dud
  is connected to the massive body, and therefore the interferometer
  ``works'', i.e. $D_2$ never clicks.)  }
\end{figure}

In order to test a bomb we have to tune the interferometer in the way
indicated above and replace one of its mirrors by the mirror-trigger of the
bomb, see Fig 2.  We send  photons through the system. If the bomb is
a dud then only detector $D_1$ clicks.
If, however the bomb is good then no interference takes place and
there are three possible outcomes: the bomb might explode (probability  ${1/2}$), detector
$D_1$ might click (probability  ${1/4}$), and detector $D_2$ might
click (probability  ${1/4}$). In the latter case we have achieved our goal:
we know that the bomb is good (otherwise $D_2$ could not click) without exploding it.

\section{The Elitzur-Vaidman bomb testing problem}

Although conceptually Penrose's proposal for testing bombs is very
interesting, its  implementation  is very difficult.
The tricky part is replacing the mirror of a tuned interferometer by
the bomb. Usually, replacing a mirror requires re-tuning the
interferometer, but here it is impossible to do that (the bomb might
explode while we are re-tuning).  The original bomb testing
problem (Elitzur and Vaidman, 1993) is slightly less dramatic, but it
can and has been implemented in a real experiment.  This problem is
equivalent to the task of finding an ultra-sensitive mine without
exploding it.  The solution is similar to the above: we tune the Mach-Zehnder
interferometer to have no photons at detector $D_2$. Then we place the
interferometer in such a way that the mine (if present) blocks one of
the arms of the interferometer, see Fig. 3.  All the discussion above
about the operation of the device is the same, but the main difficulty
is absent: there is no need to fix the  exact position of the mine.

\begin{figure} [t]
\epsfysize=4.5 cm
 \centerline{\epsfbox{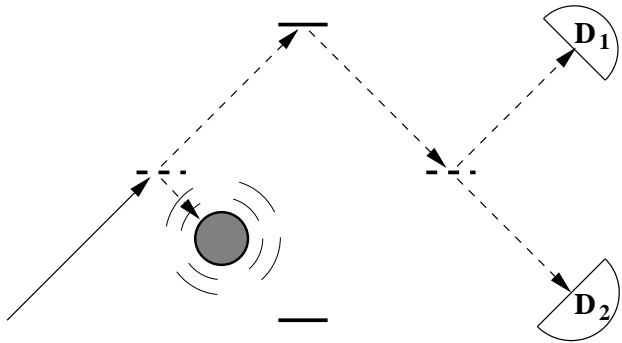}} 
\caption{{\bf Finding an ultra-sensitive mine without exploding it.}
  If the mine is present, detector $D_2$ has probability 25\% to
  detect the photon we send through the interferometer and in this
  case we know that the mine is inside the interferometer without
  exploding it.}
\end{figure}

The efficiency of finding a good bomb (mine) without exploding it in
the above procedure is only 25\%.  By modifying the transmission
coefficients of the beam splitters and repeating the procedure in case
of no explosion we can (almost) reach the efficiency of 50\%. Even more
surprisingly, the efficiency can be made as close as we want to
$100\%$ by integrating the idea of the IFM with the quantum Zeno
effect, see Kwiat et al. (1995).

\section{Experimental realization of the IFM}

Two experiments verifying IFM have been performed.  The first (Kwiat
et al.  1995) used a down-conversion crystal as a photon source and
state of the art optical components. The ``bomb'' was an efficient
detector and the ``explosion'' was a registration of the count by a
computer. A  statistical analysis of thousands of counts were performed
and, after normalization according to the noise and the efficiency
of the detectors, the results confirmed the prediction of quantum
theory with a very good precision.  The second experiment was
performed as a demonstration at the science fair in Groningen (du
Marchie van Voorthuysen, 1997).  In this experiment  a standard
Mach-Zehnder interferometer was operated. Occasionally 
a detector (connected to a loud bell in order to simulate the
explosion) was inserted along one arm of the interferometer.  A dimmed laser light entered the interferometer. The
visitors could see that without the bell-detector usually $D_1$
clicked first. When the bell-detector was present, in most of the cases
they first heard the bell, sometimes $D_1$, but also, in many cases
$D_2$ clicked first, thus telling that the bell-detector is inside the
interferometer.

The technical achievements of the two experiments cannot be compared,
but the latter had one advantage: it was really testing (although
inefficiently and not very reliably) the presence of the bomb (the
bell-detector).  The first experiment was much more precise,
but it was used more to test quantum mechanics, than to find an object
without ``touching'' it. In the experimental
run the detector which played the role of the bomb was triggered many
many times.  It  can be considered an IFM only during
the short time windows defined by the clicks of the detections of the
``idler'' photons coming from the down-conversion crystal.

\begin{figure}
  \epsfysize=5 cm \centerline{\epsfbox{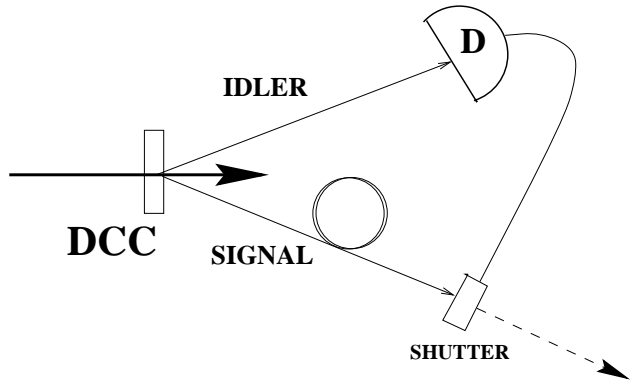}}
\caption{{\bf Single-photon gun.} The signal photon comes from the
  down-conversion crystal DCC and stored in the delay ring during the
  time that the idler photon activates a detector which sends a fast
  electronic signal to open a shutter for a short time on the way of
  the signal photon.  }
\end{figure}

 This type of experiment with coincidence counts  is commonly
considered as ``a single-photon'' experiment. (In the Groningen 
experiment there was no attempt to use a single-photon source at all.)  I think, however, that it is
conceptually important to perform an IFM with a real single-photon
source, a device which emits when commanded  just one photon.  I may call it a
single-photon gun, see Fig .4. In such an experiment there is another
paradoxical aspect: we can get information about a region of space 
never  visited by any particle, see Vaidman (1994a).

\section{Generalized IFM}
  
Let us consider now a more general task which can be considered as an
IFM, see Vaidman (1994b). We have to  verify that the
system is in a certain state, say $\vert \Psi \rangle$.
The state $\vert \Psi \rangle$ is such that its detection 
causes an explosion or destruction: destruction of a system,
of a measuring device, or at least of the state $\vert \Psi \rangle$
itself.  The states orthogonal to $\vert \Psi \rangle$ do not cause
the destruction.  Although the only physical effect of $\vert \Psi
\rangle$ is an explosion which destroys the state, we have to detect
it without any distortion. 

Let us assume that if the system is in a state $\vert \Psi \rangle$
and a part  of the measuring device is in a state $\vert \Phi_1 \rangle$, we have
an explosion.  For simplicity, we will assume that if the state of the
system is orthogonal to $\vert \Psi \rangle$ or the part of the
measuring device which interacts with the system 
is in a state $\vert \Phi_2 \rangle$ (which is orthogonal to $\vert
\Phi_1 \rangle$) than neither the system nor the measuring device
changes their state:
 \begin{equation}
  \begin{array}{clcr}
 |\Psi\rangle
~~|\Phi_1\rangle & \rightarrow  |expl \rangle  \\
 |\Psi_\perp
\rangle |\Phi_1\rangle & \rightarrow  |\Psi_\perp \rangle
|\Phi_1\rangle \\
 |\Psi\rangle ~~|\Phi_2\rangle & \rightarrow 
|\Psi\rangle ~~|\Phi_2\rangle \\
 |\Psi_\perp \rangle |\Phi_2\rangle
 & \rightarrow  |\Psi_\perp \rangle |\Phi_2\rangle .
 \end{array}
\end{equation}
Now, let us start with an initial state of the  measuring device $|\chi
\rangle = \alpha |\Phi_1\rangle + \beta |\Phi_2\rangle .$ If the
initial state of the system is $\vert \Psi \rangle$, then the
measurement interaction is:
\begin{equation}
 |\Psi \rangle |\chi \rangle \rightarrow \alpha
|expl \rangle + \beta |\Psi \rangle |\Phi_2\rangle = \alpha
|expl \rangle + \beta |\Psi \rangle (\beta^{\ast} |\chi \rangle +
\alpha |\chi_{\perp} \rangle) ,
 \end{equation}
 where $ |\chi_{\perp}
\rangle = - \beta^{\ast} |\Phi_1\rangle + \alpha |\Phi_2\rangle$.  If,
instead, the initial state of the system is orthogonal to $\vert \Psi
\rangle$, then the measurement interaction is:
 \begin{equation}
|\Psi_\perp \rangle |\chi \rangle \rightarrow |\Psi_\perp \rangle
|\chi \rangle .
 \end{equation} 
 To complete our measuring procedure we
perform a measurement on the part of the measuring device to  distinguish
between $|\chi \rangle$ and $|\chi_{\perp} \rangle$.  Since there is
no component with $|\chi_{\perp} \rangle$ in the final state (4.3), it
can be obtained only if the initial state of the system had the
component  $\vert
\Psi \rangle$.  This is also the final state of the system: we do not
obtain $|\chi_{\perp} \rangle$ in the case of the explosion.
 
A Mach-Zehnder interferometer  is a particular implementation
of this scheme. Indeed, the photon entering  the interferometer can
be considered as a measuring device prepared by the
first beam splitter in a state
$|\chi \rangle =
{1\over\sqrt 2}  (|\Phi_1\rangle  + |\Phi_2\rangle)$,
where
$|\Phi_1\rangle$ designates a photon moving  in the lower arm of the
interferometer, and
$|\Phi_2\rangle$ designates a photon moving  in the upper arm.
Detector $D_2$ together with the second beam splitter tests for the
state $|\chi_{\perp} \rangle =
{1\over\sqrt 2}  (|\Phi_1\rangle  - |\Phi_2\rangle)$.

\section{Applications of the  IFM}

In principle, the IFM may  have many dramatic
practical applications. For example, if we have a method of selecting
a certain bacteria which also kills it, the IFM may allow us to select
many such live bacteria. One can  fantasize about
 safe X-ray photography, see Fig. 5. One can design a scheme of a
computer which will allow  knowing the result of a computation  without
computing, see Jozsa (1995). These are gedanken ideas, but I am
optimistic about finding situations in which the IFM will have some real
practical applications.

\begin{figure}
\epsfysize=5 cm
 \centerline{\epsfbox{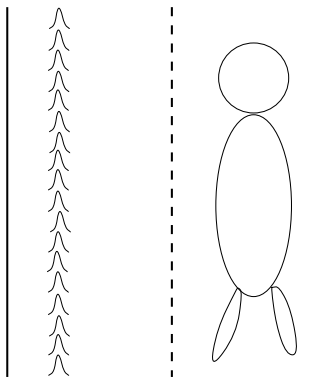}} 
\caption{{\bf Safe X-ray photography.} Between two parallel mirrors
  another mirror which reflects 99.9\% and transmits 0.1\% is
  introduced.  A photon starting on the left side will then end up on
  the right side after the about 50 bounces. If, however, on the right
  side there is an object which can absorb the photon, then with 95\%
  probability the photon will stay on the left.  To make a
  photographic picture, a person enters into the right side and the
  photons are placed on the left side. After the time required for 50
  bounces a photographic plate is introduced in the left side. In such
  an experiment the bones absorb only 5\% of the radiation, i.e.
  twenty times less than in the the standard method. (It is assumed
  that the soft tissue does not affect the X-rays at all, while the
  bones are opaque.)}
\end{figure}

\section{The IFM as counterfactuals}

 The IFM  also has interesting philosophical implications. Let me
quote  
Roger Penrose (1994, p.240):
\begin{quotation}
  \noindent
  What is particularly curious about quantum theory is that there can
  be actual physical effects arising from what philosophers refer to
  as {\em counterfactuals} -- that is, things that might have
  happened, although they did not happened. 
\end{quotation}
For me this paradoxical situation gives another reason why we should
accept the Many-Worlds Interpretation (MWI) of quantum theory.
According to the MWI, in the situations considered by Penrose,
``things'' did not happened in a particular world, but did happened in
some other world (see Vaidman 1994a). Therefore, they did take place
in the physical universe which incorporates all the worlds and thus
their effect is not so surprising.

 The research was supported in part by
grant
614/95 of the the Basic Research Foundation (administered by the
Israel Academy of Sciences and Humanities).

 \end{document}